\documentclass[twocolumn]{aastex631}

\usepackage{amsmath} 

\renewcommand{\textbf}[1]{#1}  

\begin{document}

\title{High-Velocity Molecular Clouds in M83}

\author[0009-0004-9830-4543]{Maki Nagata}
\affiliation{Institute of Astronomy, Graduate School of Science, The University of Tokyo \\
2-21-1 Osawa, Mitaka, Tokyo 181-0015, Japan}

\author[0000-0002-1639-1515]{Fumi Egusa}
\affiliation{Institute of Astronomy, Graduate School of Science, The University of Tokyo \\
2-21-1 Osawa, Mitaka, Tokyo 181-0015, Japan}

\author[0000-0002-8868-1255]{Fumiya Maeda}

\affiliation{Research Center for Physics and Mathematics, Osaka Electro-Communication University\\
18-8 Hatsucho, Neyagawa, Osaka 572-8530 Japan}

\author[0000-0002-2062-1600]{Kazuki Tokuda}
\affiliation{Faculty of Education, Kagawa University, Saiwai-cho 1-1, Takamatsu, Kagawa 760-8522, Japan}
\affiliation{Department of Earth and Planetary Sciences, Faculty of Science, Kyushu University, \\Nishi-ku, Fukuoka, Fukuoka 819-0395, Japan}
\affiliation{National Astronomical Observatory of Japan, National Institutes of Natural Sciences \\2-21-1 Osawa, Mitaka, Tokyo 181-8588, Japan}

\author[0000-0002-4052-2394]{Kotaro Kohno}
\affiliation{Institute of Astronomy, Graduate School of Science, The University of Tokyo \\
2-21-1 Osawa, Mitaka, Tokyo 181-0015, Japan}

\author[0000-0003-3932-0952]{Kana Morokuma-Matsui}
\affiliation{Institute of Astronomy, Graduate School of Science, The University of Tokyo \\
2-21-1 Osawa, Mitaka, Tokyo 181-0015, Japan}

\author[0000-0002-8762-7863]{Jin Koda}
\affiliation{Department of Physics and Astronomy, Stony Brook University \\ Stony Brook, NY 11794-3800, USA}




\begin{abstract}
   \vskip.5\baselineskip
   \textcolor{black}{High-velocity clouds (HVCs), which are gas clouds moving at high velocity relative to the galactic disk, may play a critical role in galaxy evolution, potentially supplying gas to the disk and triggering star formation.
   In this study, we focus on the nearby face-on barred spiral galaxy M83, where high spatial resolution, high-sensitivity CO (1--0) data are available. We identified molecular clouds
   and searched for clouds with velocities deviating by more than 50\,km\,$\text{s}^{-1}$ from the disk velocity field as HVCs.}
   \textcolor{black}{A total of 10 HVCs were detected—nine redshifted and one blueshifted—clearly highlighting an asymmetry in their velocity distribution.} These HVCs  have radii of 30--80 pc, masses on the order of $10^5$\,$M_{\odot}$, and velocity dispersions of 3--20 km\,$\text{s}^{-1}$, displaying a tendency toward higher velocity dispersion compared to disk molecular clouds in M83. Most of the HVCs do not overlap with the candidates of supernova remnants, and the energy needed to drive HVCs at such high velocities exceeds single supernova energy. Together with the asymmetry in their velocity distribution, we thus conclude that most of the HVCs found in this study are inflow from outside the M83’s disk.
\end{abstract}



\section{Introduction} \label{sec:intro}
\subsection{High-Velocity Clouds}
Disk galaxies in the local Universe are rotation dominated in general. \textcolor{black}{However, the interstellar gas in galaxies is exposed to highly dynamic environments influenced by various processes. These include feedback from massive stars, such as supernova explosions and stellar winds, galactic fountains powered by jets from massive black holes (active galactic nuclei; AGNs), and the inflow of gas from external sources. (see the review paper \citet{Mac_Low2004} for stellar feedback, \citet{Veilleux2005} for AGN fountain and \citet{Sancisi2008} for gas inflow.)} Some of this gas is known to move at peculiar velocities, differing from the galactic rotation by tens to hundreds of km\,$\text{s}^{-1}$. Such anomalous-velocity gas impacts the lifecycle of interstellar matter and galaxy evolution through mechanisms such as star formation triggered by cloud-cloud collisions \textcolor{black}{(see the review paper by \citet{Fukui2021})}, gas recycling in galactic fountains, and the redistribution of matter via gas inflow into the galactic disk and outflow into intergalactic space. Therefore, investigating the distribution of these high-velocity gas clouds and elucidating their origins are the key to understanding star formation and galactic evolution.

In the Milky Way (MW), high-velocity clouds (HVCs) with velocities exceeding approximately 50 $\text{km}\,\text{s}^{-1}$ relative to Galactic rotation have been primarily detected via emission lines from neutral atomic hydrogen \citep{Wakker1991}. These HVCs are estimated to be located within 10 kpc from the thick Galactic disk, with masses on the order of $10^5-10^6\,\,M_\odot$. Their origins are thought to be associated with Galactic fountain, gas accretion from satellite galaxies or the cooling of overdense regions in the multi-phase halo medium \citep{Putman_2011, Putman2012}. Some HVCs have also been detected in ultraviolet absorption lines of molecular hydrogen and infrared dust continuum emission \citep{Gillmon2006,Wakker2006,Miville2005}, though molecular gas emission lines (e.g., CO) have largely remained undetected \citep{Dessauges2007}, with the exception of CO (2--1) emission detected in two HVCs flowing out from the Galactic center \citep{DiTeodoro2020}, \textcolor{black}{which are estimated to have masses of several hundred solar masses.} While such observations exist, since we observe the Galactic disk edge-on, it is challenging to comprehensively and quantitatively measure the properties of these HVCs, such as distance, and investigate their relationship with the Galactic structure. Thus, spatially resolved observations of high-velocity gas clouds in nearby face-on galaxies are crucial. \textcolor{black}{In this study, we focus on the barred spiral galaxy M83 and investigate the HVCs within it.}

\subsection{M83}
M83 is a barred spiral galaxy located at a distance of 4.5 Mpc \citep{Thim2003}. With an inclination angle of \(26^{\circ}\) \citep{Koda2023}, it is nearly face-on, making it highly suitable for this study. The total stellar mass of M83 is \(M_{\text{star}}=2.5\times10^{10} \, M_{\odot}\) \citep{Jarrett2019}, and its molecular gas mass is \(M_{\text{H}_2}=2.6\times10^{9} \, M_{\odot}\) \citep{Koda2023}. In comparison, the MW has a stellar mass of \(M_{\text{star}}=4.6\times10^{10} \, M_{\odot}\) \citep{Bovy2013} and a molecular gas mass of \(M_{\text{H}_2}=1.0\times10^{9} \, M_{\odot}\) \citep{Chomiuk2011}. \textcolor{black}{Both galaxies share similar characteristics in terms of stellar and molecular gas masses, and their barred spiral morphology further highlights their resemblance.} Consequently, M83 has been observed across multiple wavelengths, including broadband and narrowband imaging with the Hubble Space Telescope (HST) \citep{Blair2014}, three-dimensional spectroscopic observations with the Multi Unit Spectroscopic Explorer (MUSE) \citep{Della2022}, and H$\;${\sc i} observations with the Very Large Array (VLA) \citep{Eibensteiner2023}. Additionally, catalogs of supernova remnants \citep{Long2022} are available, enabling studies from multiple perspectives.



\textbf{\citet{Miller2009} conducted a H$\;${\sc i} survey of M83 and identified eight HVCs with the velocities $40$--$170\,\text{km} \,\text{s}^{-1}$ offset from the disk velocity field. These clouds span an H$\;${\sc i} mass range of $7 \times 10^5$ to $1.5 \times 10^7 \, M_\odot$, \textcolor{black}{with three of them located in the disk region analyzed in this work and the remaining five in outside the disk.}}
\textbf{\textcolor{black}{A more recent study by \citet{Della2022} reported a unique arc-like feature (Feature C) based on MUSE H$\alpha$ data.}} The feature shows a velocity offset of $\sim30 \, \text{km} \, \text{s}^{-1}$ and a velocity dispersion of $\sim80 \, \text{km} \, \text{s}^{-1}$, and is located approximately 0.8 kpc east of the galactic center.
In this region, elevated [N$\;${\sc ii}]/H$\alpha$ and [O$\;${\sc iii}]/H$\alpha$ ratios indicated the presence of shocks on the [N$\;${\sc ii}]–based Baldwin–Phillips–Terlevich (BPT) diagram. \textcolor{black}{This region was also investigated using CO(2--1) data from the Atacama Large Millimeter/submillimeter Array (ALMA), which revealed a velocity offset of $\sim100\,\text{km} \, \text{s}^{-1}$ from the disk and a velocity dispersion of $\sim25 \, \text{km} \, \text{s}^{-1}$.}

\textbf{In this study, we use high-sensitivity CO(1–0) mapping data from \citet{Koda2023}, which cover nearly the entire disk of M83, to investigate the nature of HVCs with anomalous velocities perpendicular to the disk. This provides a complementary view to the previous H$\;${\sc i} and CO(2–1) studies.}

\section{Method}
\subsection{Observation Data}
The CO(1--0) data observed with the ALMA telescope \citep{Koda2023} were used. The spatial resolution is $2\farcs12\times\ 1\farcs71$ at a position angle of $-89^{\circ}$, corresponding to a physical scale of $\sim$40 pc. The velocity resolution is \(5 \, \text{km} \, \text{s}^{-1}\), with a sensitivity ($\sigma$) of 3.2 mJy\,beam$^{-1}$ (80 mK ; calculated from a distance of 4.2$^\prime$ from the galactic center in non-detection channels), allowing for high-sensitivity molecular cloud detection \textcolor{black}{(With a mass sensitivity of $10^4 M_{\odot}$, as explained in the next section)}. The field of view (FoV) spans a diameter of 9.4$^\prime$ ($\sim$12.3 kpc), providing extensive spatial coverage for detailed analysis. \textcolor{black}{Each pixel corresponds to $0\farcs25\times 0\farcs25$, ensuring detailed sampling of the observations.}

\subsection{Method for Detecting HVCs}

\textcolor{black}{First, we used the \texttt{astrodendro} software package \citep{Astrodendro2019} to identify virtually all molecular clouds in the data cube,} setting the threshold parameters as follows: 
\textcolor{black}{\texttt{min\_value} was set to 2\,$\sigma$, \texttt{min\_delta} to 3.5\,$\sigma$, and \texttt{min\_npix} to 200 voxels, equivalent to about four times the beam area.}
\textcolor{black}{A cloud would be identified based on this criterion if it has a peak temperature of at least 5.5\,$\sigma$ and contains more than 200 voxels with intensities above 2\,$\sigma$, corresponding to a minimum mass range of \(\sim2.1 \times 10^4 \, M_\odot\) (for 200 voxels at 2\,$\sigma$) to \(\sim5.7 \times 10^4 \, M_\odot\)(for 200 voxels at 5.5\,$\sigma$)} \textcolor{black}{assuming the MW CO-to-$\mathrm{H_2}$ conversion factor (see Section \ref{formula}).}
Among the identified structures, only the \textit{leaf} structures were considered molecular clouds. 

\textcolor{black}{Note that we do not use the cloud catalog by \citet{Hirota2024}, which is based on a 1\,km\,s$^{-1}$ resolution cube, but instead re-identified clouds using the 5\,km\,s$^{-1}$ resolution cube as described above. This is because the catalog based on the 1\,km\,s$^{-1}$ resolution data cube was found to artificially split the HVCs with a wide velocity dispersions into multiple components.}


In order to measure the relative velocity of molecular clouds to the galactic disk, we derived the large-scale velocity field \textcolor{black}{(hereafter also referred to as the disk velocity field, $v_\text{disk}$)} \textcolor{black}{by using the \texttt{imsmooth} task in \textsc{CASA} \citep{McMullin2007,CASA2022} }to smooth the cube to 400 pc and then calculating the intensity-weighted velocity. The smoothing scale of 400 pc was chosen as it approximately matches the arm thickness, helping to remove local flows associated with the arm. \textcolor{black}{Gas dynamics in the central region of M83 are complicated due to circumnuclear ring \citep{Elmegreen1998}, and thus the velocity field from the 400-pc-smoothed data cube does not represent true disk velocities in this region. We therefore exclude molecular clouds at $R_\text{galcen}<0.4$ kpc from the following analysis, and compute the velocity deviation $\Delta v$ \textcolor{black}{($\Delta v$ $\equiv v_\text{LSR}^\text{HVC} - v_\text{disk}$ ; $ v_\text{LSR}^\text{HVC}$ is the mean velocity of the
cloud from the \texttt{v\_cen} parameter in  \texttt{astrodendro} output)} of molecular clouds  located at 0.4 kpc $<R_\text{galcen}<5.5$ kpc by subtracting the disk velocity field from the center velocity of the clouds.}
We also excluded identified clouds whose distance from the galactic center $R_\text{galcen}$ exceeds 5.5 kpc \textcolor{black}{as the noise level increases around the border of FoV. }

\textcolor{black}{In this study, we focus on HVCs moving perpendicular to the galactic plane and exclude non-circular motions within the galactic disk. Since streaming motions due to spiral arms are typically $\sim10 \,\text{km}\,\text{s}^{-1}$ \citep{Baba2016}, }\textcolor{black}{and those observed within the bars reach $\sim30-40\,\text{km}\,\text{s}^{-1}$ along the line of sight in M83 \citep{Hirota2014}, we define HVCs as clouds with absolute velocity deviations $|\Delta v|$ of $50\,\text{km}\,\text{s}^{-1}$ or more from the disk velocity field. }


\subsection{Calculation of the Properties of HVCs}\label{formula}


\textbf{\textcolor{black}{To compute the molecular gas mass (including helium and heavier elements), we adopted the Galactic CO-to-$\text{H}_2$ conversion factor $\alpha_{\text{CO}} = 4.3 \,M_\odot\,(\mathrm{K\,km\,s^{-1}\,pc^{-2}})^{-1}$ \citep{Bolatto2013}, as given by the following equation.
\begin{align}
    M_{\text{mol}} &=\alpha_{\text{CO}}L_{\rm CO} ,
\end{align}
where $L_{\rm CO}$ is the CO luminosity for each identified structure. (Note that the actual conversion factor in M83 may be lower in the inner $\sim3\,\rm kpc$ \citep{Lee2024}, as discussed in Section~\ref{uncertainties}.
)}}

\textcolor{black}{\texttt{Astrodendro} outputs the geometric mean of the \texttt{major\_sigma} ($\sigma_{maj}$) and \texttt{minor\_sigma} ($\sigma_{min}$) as effective rms size. As suggested by \citet{Solomon1987}, we multiplied the output  effective rms size by 1.91 to obtain the molecular cloud radius ($r=1.91\times\sqrt{\sigma_{maj}\times\sigma_{min}}$)}, \textcolor{black}{which corresponds to the FWHM and is reported as 'radius' in Table \ref{table}.}
\textcolor{black}{\texttt{Astrodendro} also outputs the intensity-weighted second moment of velocity \texttt{v\_rms} ($\sigma_v$ in Table \ref{table}).}

\textbf{Following \citet{Mckee1992}, the virial mass ($M_{\rm vir}$ in Table \ref{table}) is estimated as
\begin{equation}
M_{\rm vir} = \frac{5r\sigma_v^2}{G}.
\end{equation}
The virial parameter is defined as
\begin{equation}
\alpha_{\rm vir} = \frac{M_{\rm vir}}{M_{\rm mol}}.
\end{equation}
The virial parameter is a measure of the ratio of kinetic to gravitational energy, and serves as an indicator of whether a cloud is gravitationally bound.}






\section{Results}

\begin{figure*}[htbp]
    \centering
    \includegraphics[width=0.95\textwidth]{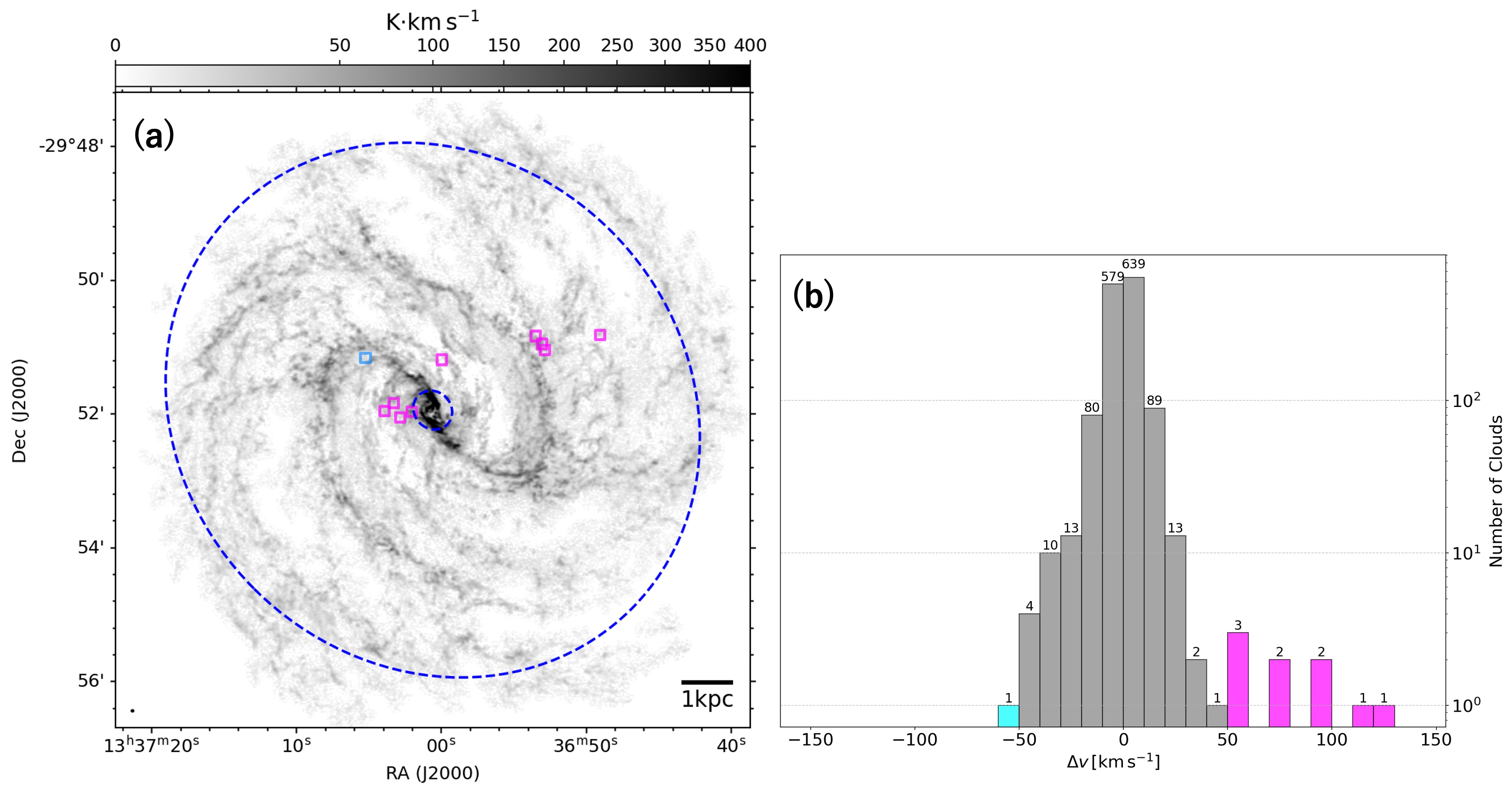}
    \caption{(a) Distribution of HVCs (magenta squares represent clouds with a velocity difference of +50 km\,$\text{s}^{-1}$ or more, while the cyan square indicates a cloud with a velocity difference of $-50  \text{km}\,\text{s}^{-1}$ or less). Blue dashed circles show galactocentric radii of 0.4 kpc and 5.5 kpc. The background image is an CO(1--0) integrated intensity map created by \citet{Koda2023}. \textcolor{black}{(b) Distribution of the number of clouds in each $\Delta v$ bin.}}
    \label{mom0_hvc}
\end{figure*}

\subsection{Properties of HVCs}
\textcolor{black}{A total of $\sim$1400 molecular clouds have been identified within 0.4 kpc $< R_{\text{galcen}} <$ 5.5 kpc. They have radii $r$ of $~$20 – 120 pc, masses $M_{\rm mol}$ in the range of $10^4$ – $10^7 \ M_{\odot}$, and velocity dispersions $\sigma_v$ of $\sim2$ – 20 km s$^{-1}$. The radii and masses of the clouds identified in this study are largely consistent with those cataloged by \citet{Hirota2024} \textcolor{black}{who used a 1 km s$^{-1}$ resolution cube instead of the 5 km s$^{-1}$ cube used in this study}, except that smaller clouds in terms of radius and mass are not detected. Additionally, compared to the catalog of \citet{Hirota2024}, the velocity dispersions tend to be larger, likely due to the difference in velocity resolution between the datasets.}

\textcolor{black}{Among the $\sim$1400 identified clouds, 10 are classified as HVCs, accounting for  $\sim$1\% of the total. Figure \ref{mom0_hvc}(b) presents the distribution of cloud numbers as a function of velocity deviation $\Delta v$, binned in 10 km s$^{-1}$ intervals.} \textcolor{black}{The figure clearly shows that most clouds have velocity deviations of $|\Delta v| < 40$ km s$^{-1}$. Therefore, the definition of HVCs as those with $|\Delta v| > 50$ km s$^{-1}$ effectively identifies kinematically peculiar objects.} \textcolor{black}{Among the HVCs, nine are found on the redshifted side relative to the disk, while only one is on the blueshifted side, showing an asymmetry in the velocity distribution. }

\begin{figure*}[htbp]
    \centering
    \includegraphics[width=0.8\textwidth]{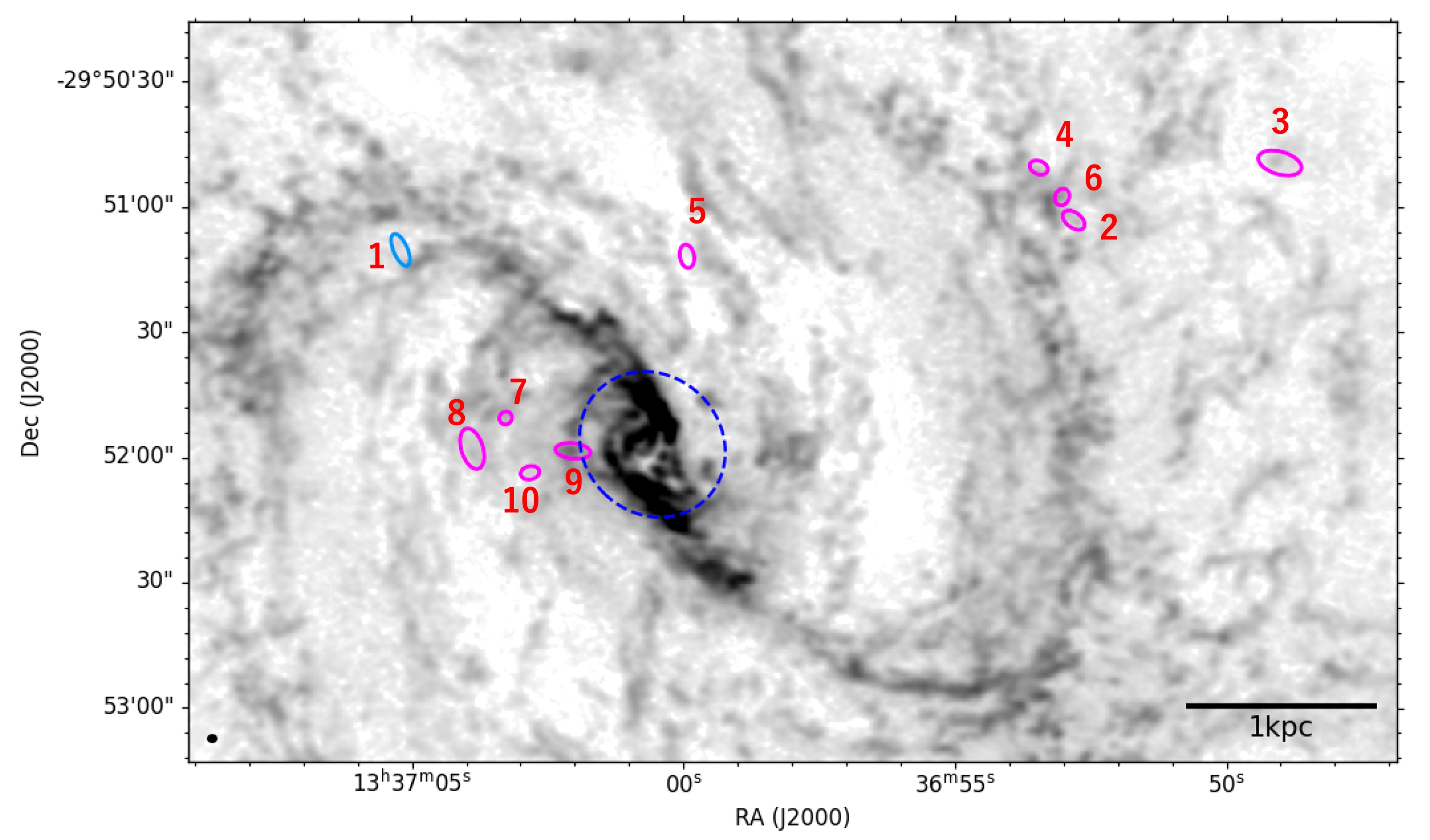}
    \caption{Zoomed-in map highlighting the locations of HVCs. Magenta ellipses represent clouds with a velocity difference of +50 km\,$\text{s}^{-1}$ or more, while the cyan ellipse shows a cloud with a velocity difference of $-$50 km\,$\text{s}^{-1}$ or less. The major and minor axes of the ellipses are derived from the major and minor standard deviations (\texttt{major\_sigma} and \texttt{minor\_sigma}) obtained from the \texttt{astrodendro} output. These values are multiplied by 1.91 to convert them into FWHM \citep{Solomon1987}. The red numbers correspond to the HVC IDs listed in Table \ref{table}. The blue dashed circle shows galactocentric radius of 0.4 kpc. 
    }
    \label{number}
\end{figure*}

The spatial distribution of HVCs is shown in Figure \ref{mom0_hvc}(a).  Although HVCs are primarily found near the center, some are found in regions somewhat offset from the arm structure, \textcolor{black}{ where streaming motions are not particularly strong. HVC 1 (the blue square in Figure \ref{mom0_hvc}(a)) is positioned at the bar end where the non-circular motions are generally strong. However, according to \citet{Hirota2014}, the streaming motion of the bar at this location acts toward the redshifted direction, while HVC 1 shows a blueshift relative to the disk velocity field. We thus regard that HVC 1 is not bar-driven.}
Figure \ref{number} presents a zoomed-in map highlighting the locations and shapes of HVCs. The major and minor axes of the ellipses are derived from the major and minor standard deviations (\texttt{major\_sigma} and \texttt{minor\_sigma}) obtained from the \texttt{astrodendro} output. These values are multiplied by 1.91 to convert them into FWHM \citep{Solomon1987}.

\textcolor{black}{\textbf{HVCs 2, 3, 4, and 6 spatially overlap with one of the H$\;${\sc i} HVCs identified by \citet{Miller2009}, designated as AVC 1 (AVC stands for anomalous-velocity clump). AVC 1 shows a velocity offset of +51 km\,s$^{-1}$ relative to the disk, and our CO-detected HVCs in this region exhibit velocity offsets of 56–72 km\,s$^{-1}$, suggesting a spatial and kinematic association.}
\textbf{In \citet{Della2022}, the arc-like feature (Feature C) with a velocity offset of $\sim$30\,km\,s$^{-1}$ in H$\alpha$ was reported, which spatially aligns with HVCs 7 and 8. They also identified a high-velocity component in CO(2--1) corresponding to HVC 8.}}

Table \ref{table} summarizes the properties of all identified HVCs. \textcolor{black}{These HVCs  have radii of 30--80 pc, masses on the order of $10^5$\,$M_{\odot}$, and velocity dispersions of 3--20 km\,$\text{s}^{-1}$.} 
Their median mass ($1.8\times10^5\,M_{\odot}$) is slightly smaller compared to the median mass of disk clouds  ($2.7\times10^5\,M_{\odot}$). Their scaling relations are discussed in Sec. \ref{scaling relation}.

\begin{figure*}[htbp]
    \centering
    \includegraphics[width=0.9\textwidth]{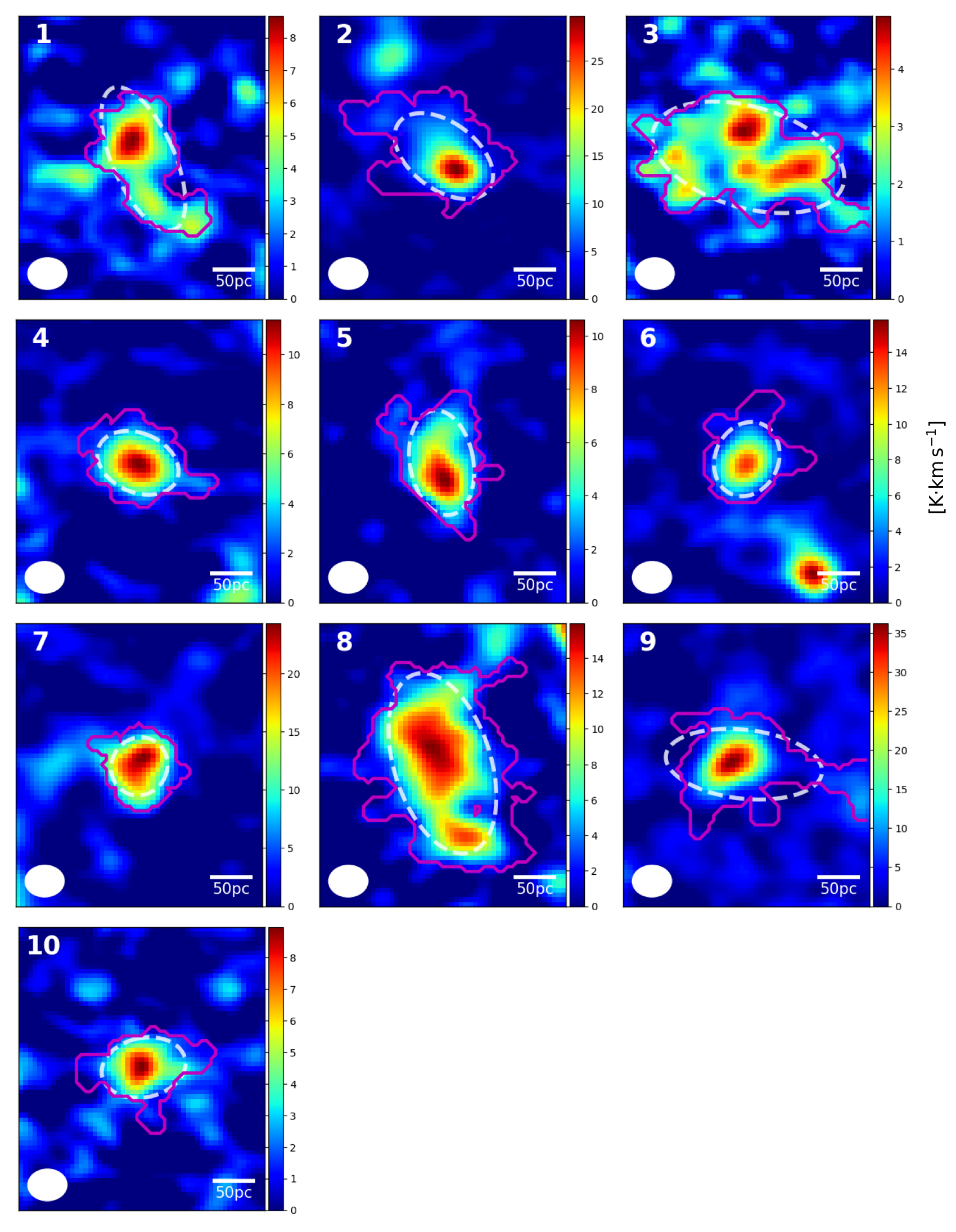}
    \caption{Integrated intensity map of HVCs: \textcolor{black}{the velocity channels were integrated over a width of $4\sigma_v$, centered at the \texttt{v\_cen}}. \textcolor{black}{Magenta contours outline the region of each HVC identified by \texttt{astrodendro}. The major and minor axes of the white dashed ellipses correspond to \texttt{major\_sigma} and \texttt{minor\_sigma} (from \texttt{astrodendro}) multiplied by 1.91 (i.e. same as the ellipses in Figure \ref{number}). \textbf{The angular resolution is indicated by the ellipses in the bottom left.}}}
    \label{mom0_panel}
\end{figure*}

\begin{figure*}[htbp]
    \centering
    \includegraphics[width=0.8\textwidth]{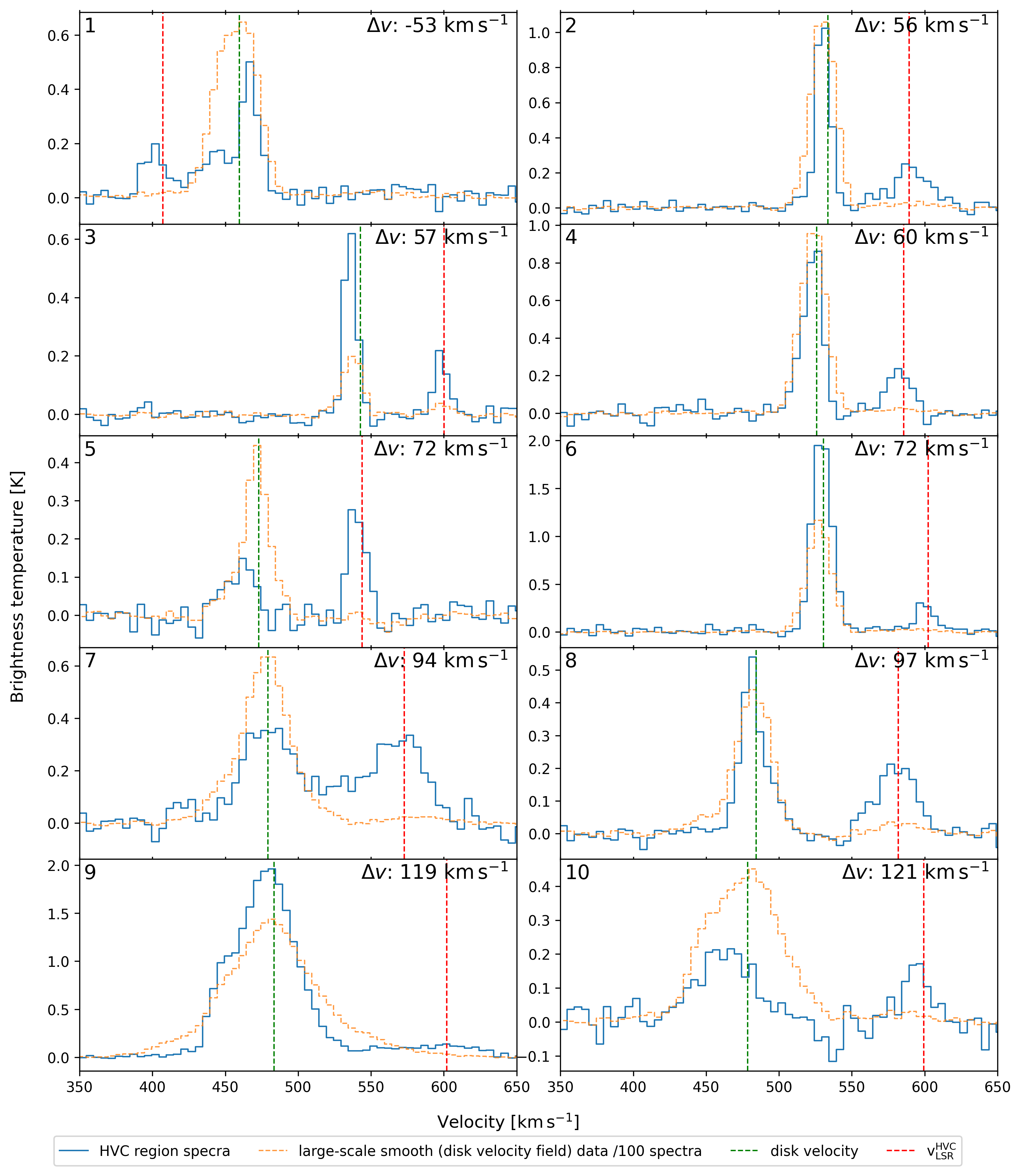}
    \caption{The spectra of HVCs: blue lines show spectra averaged over each HVC's identified region. Orange dashed lines show the spectra of the large-scale (disk) velocity field (divided by 100) within the same region. Green dashed lines mark the disk velocity at the cloud center. Red dashed lines represent $v_\text{LSR}^{\rm HVC}$ of HVCs. $\Delta v$, displayed in the top right corner of each panel, represents the velocity deviation.}
    \label{spectral_panel}
\end{figure*}

Figures \ref{mom0_panel} shows the integrated intensity map of each HVC.
The central position of each HVC was set to the center of the map, and the velocity channels were integrated over a width of each $4\sigma_v$, centered at the \texttt{v\_cen}.
The spectra of HVCs are shown in Figure \ref{spectral_panel}. The blue lines represent spectra averaged over each HVC’s identified region. The spectra within these HVCs confirm a double-peak structure in the line-of-sight velocity, consisting of both the disk velocity field and the HVC component. \textbf{(See also Figure~\ref{PV} in Appendix~\ref{appendix_PV} for Position-Velocity diagrams.)}


\subsection{Scaling Relation}\label{scaling relation}

\begin{figure*}[htbp]
    \centering
    \includegraphics[width=0.97\textwidth]
    {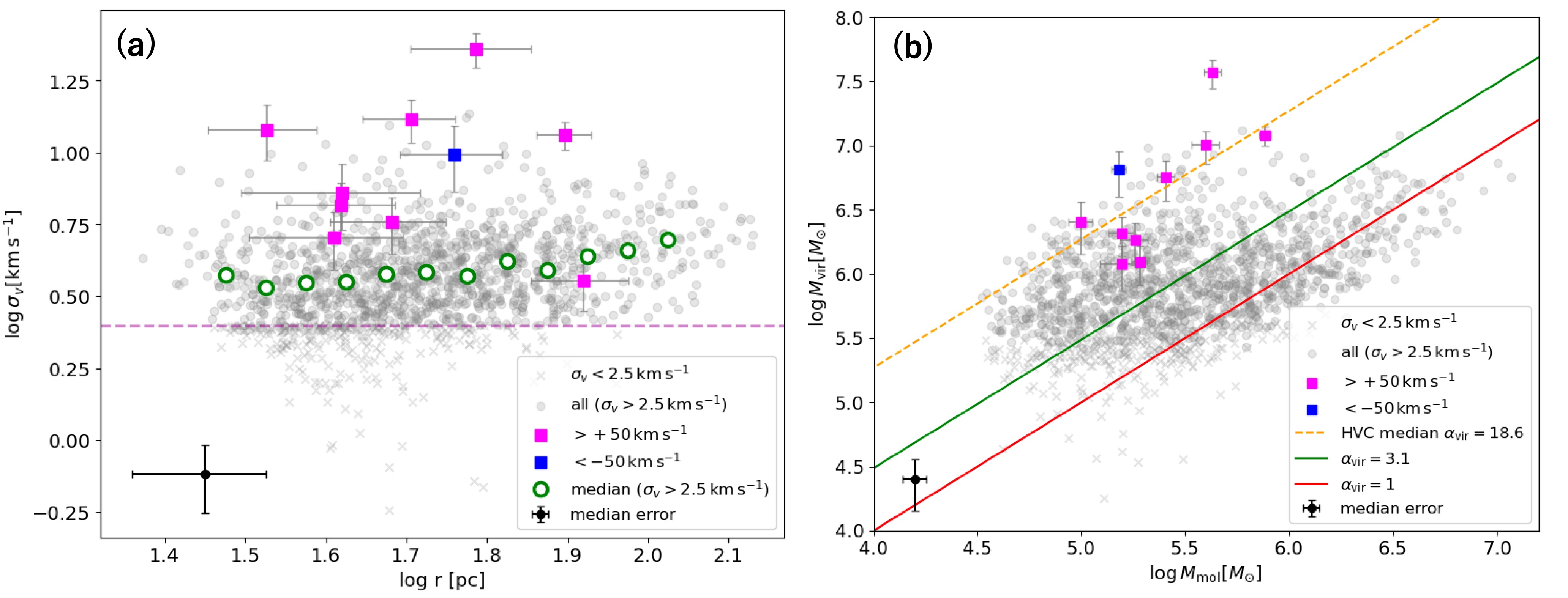}
    \caption{Scaling relations of cloud properties. In each panel, magenta squares indicate molecular clouds with $\Delta v > +50\,\text{km}\,\text{s}^{-1}$, the blue square indicates a cloud with $\Delta v < -50\,\text{km}\,\text{s}^{-1}$, and gray dots represent disk clouds. Crosses denote clouds with $\sigma_{v} < 2.5\,\text{km}\,\text{s}^{-1}$. \textbf{Error bars represent statistical uncertainties derived from bootstrap resampling. A representative median error for clouds with $\sigma_{v} \geq 2.5\,\text{km}\,\text{s}^{-1}$ is shown in the lower-left corner of each panel.} (a) $r$--$\sigma_v$ relation: Green dots show the median values in each bin, including HVCs but excluding clouds with $\sigma_{v} < 2.5\,\text{km}\,\text{s}^{-1}$. The purple dashed line indicates the velocity dispersion resolution limit of $2.5\,\text{km}\,\text{s}^{-1}$. \textbf{(b) $M_{\rm mol}$--$M_{\rm vir}$ relation: The red solid line corresponds to a virial parameter $\alpha_{\rm vir} = 1$. The green solid line indicates the median virial parameter of the disk clouds (excluding $\sigma_{v} < 2.5\,\text{km}\,\text{s}^{-1}$) with masses on the order of $10^5 M_\odot$, $\alpha_{\rm vir} = 3.1$. The orange dashed line shows the median $\alpha_{\rm vir} = 18.6$ for the HVCs.}}
    \label{radius_sigv}
    
\end{figure*}

\textcolor{black}{We compared HVCs with disk molecular clouds in M83 based on the relationship between radius and velocity dispersion $\sigma_{v}$ \citep{Larson1969} in Figure \ref{radius_sigv}(a).} The figure also displays the median $\sigma_{v}$ values for clouds binned by radius.
When the channel width of the velocity is taken as the FWHM, velocity dispersions smaller than approximately half of the channel width are unresolved. These data points, marked with crosses, were excluded from the median calculation. The figure shows that HVCs tend to have larger velocity dispersions, $\sim$ 0.5 dex higher than those of disk molecular clouds. This trend will be discussed in Section \ref{discussion}.

\textbf{We next plotted the relationship between the CO luminosity based gas mass and the virial mass in Figure~\ref{radius_sigv}(b). The HVCs are distributed toward the upper-left side compared to the disk clouds. \textcolor{black}{The median virial parameter of disk clouds with masses comparable to those of the HVCs (i.e., on the order of $10^5\,M_\odot$) is $\alpha_{\rm vir} = 3.1$}, whereas that of the HVCs is significantly higher  ($\alpha_{\rm vir} \sim 19$). This large value is due to their relatively high velocity dispersions, as shown in Figure~\ref{radius_sigv}(a), indicating that the HVCs are much less gravitationally bound than the disk clouds.}
\textbf{We show some other scaling relations ($M_{\rm mol}$--$\sigma_v$ and $r$--$M_{\rm mol}$) in Figure \ref{Other_scaling_relation} in Appendix \ref{other_scaling}.}

\begin{table*}[htbp]
\caption{The properties of high velocity molecular clouds (sorted by $\Delta v$)}
\centering
\begin{tabular}{c c c c c c c c c c}
\hline\hline
\textbf{ID} & R.A. & Decl. & $\Delta v$ & $v_\text{LSR}^{\rm HVC}$ & $r$ & $\sigma_{v}$ & $M_{\rm mol}$ & $M_{\rm vir}$ & $R_{\text{galcen}}$ \\
\textbf{(1)} & (2) & (3) & (4) & (5) & (6) & (7) & (8) & (9) & (10) \\
 &  & & (km\,$\text{s}^{-1}$) & (km\,$\text{s}^{-1}$) & (pc) & (km\,$\text{s}^{-1}$) & ($10^5\,M_{\odot}$) & ($10^6\,M_{\odot}$) & (kpc) \\
\hline
\textbf{1} & 13:37:05.207 & -29:51:10.32 & -52.8 & 407.1 & $57.5 \pm 8.4$ & $9.8 \pm 2.5$ & $1.5 \pm 0.1$ & $6.5 \pm 2.5$ & 1.668 \\
\textbf{2} & 13:36:52.814 & -29:51:03.17 &  56.1 & 589.3 & $50.9 \pm 6.7$ & $13.0 \pm 2.3$ & $4.0 \pm 0.6$ & $10.0 \pm 2.8$ & 2.749 \\
\textbf{3} & 13:36:49.018 & -29:50:49.47 &  57.1 & 600.0 & $83.1 \pm 11.5$ & $3.6 \pm 0.8$ & $1.9 \pm 0.2$ & $1.2 \pm 0.4$ & 3.949 \\
\textbf{4} & 13:36:53.458 & -29:50:50.58 &  59.9 & 585.6 & $41.5 \pm 7.0$ & $6.5 \pm 1.3$ & $1.6 \pm 0.2$ & $2.1 \pm 0.7$ & 2.758 \\
\textbf{5} & 13:36:59.934 & -29:51:11.83 &  71.5 & 544.0 & $48.1 \pm 7.9$ & $5.7 \pm 1.3$ & $1.8 \pm 0.3$ & $1.8 \pm 0.7$ & 1.080 \\
\textbf{6} & 13:36:53.029 & -29:50:57.63 &  71.9 & 602.3 & $40.8 \pm 8.8$ & $5.0 \pm 1.1$ & $1.6 \pm 0.4$ & $1.2 \pm 0.5$ & 2.765 \\
\textbf{7} & 13:37:03.272 & -29:51:50.57 &  93.6 & 572.8 & $33.6 \pm 5.2$ & $12.0 \pm 2.6$ & $2.6 \pm 0.2$ & $5.6 \pm 1.9$ & 0.809 \\
\textbf{8} & 13:37:03.890 & -29:51:57.91 &  96.5 & 581.8 & $79.0 \pm 6.2$ & $11.5 \pm 1.2$ & $7.7 \pm 0.5$ & $12.1 \pm 2.1$ & 0.998 \\
\textbf{9} & 13:37:02.034 & -29:51:58.46 & 118.7 & 602.1 & $61.1 \pm 10.4$ & $22.9 \pm 3.0$ & $4.3 \pm 0.4$ & $37.2 \pm 9.5$ & 0.441 \\
\textbf{10} & 13:37:02.824 & -29:52:03.68 & 120.6 & 599.3 & $41.7 \pm 10.4$ & $7.2 \pm 1.8$ & $1.0 \pm 0.1$ & $2.5 \pm 1.1$ & 0.712 \\
\hline
\end{tabular}
\label{table}
\tablecomments{
Parameters of the identified HVCs, sorted by $\Delta v$. 
(1) ID. 
(2) Right ascension (mean position of the cloud). 
(3) Declination (mean position of the cloud). 
(4) Velocity deviation. 
(5) Centroid velocity. 
(6) Radius. 
(7) Velocity dispersion. 
(8) Molecular gas mass in units of $10^5\,M_{\odot}$. 
(9) \textbf{Virial mass in units of $10^6\,M_{\odot}$}. 
(10) Galactocentric radius.
}
\end{table*}

\subsection{Uncertainties}\label{uncertainties}
\textbf{We consider two main sources of uncertainty in the derived cloud properties: statistical errors arising from measurement noise, and systematic uncertainties such as the effects of finite resolution and the choice of the $\alpha_{\rm CO}$ conversion factor.}

\textbf{We estimated the statistical uncertainties of cloud properties ($r$, $\sigma_v$, $M_{\rm mol}$, $M_{\rm vir}$) using bootstrap resampling \citep{Rosolowsky2006}. We performed 1000 bootstrap iterations with the Python package \texttt{dendroplot} \citep{dendroplot, Wong2017} \textcolor{black}{and adopted the $\pm1\sigma$ range centered on the median of the bootstrap resamplings as the measurement uncertainty.}
The resulting uncertainties are listed in Table~\ref{table}. In addition, Figure~\ref{radius_sigv} shows the error bars for individual HVCs, and the median uncertainty for clouds with $\sigma_{v} \geq 2.5\,\text{km}\,\text{s}^{-1}$ on the bottom-left corner.}

\textbf{The finite spectral resolution of the data leads to artificial broadening of the observed line widths. If no deconvolution is applied, the measured velocity dispersion $\sigma_v$ will be overestimated relative to the intrinsic dispersion $\sigma_{v, \mathrm{int}}$, following the relation
\begin{equation}
    \sigma_v = \sqrt{\sigma_{v, \mathrm{int}}^2 + \sigma_{\mathrm{ch}}^2},
\end{equation}
where $\sigma_{\mathrm{ch}}$ is the dispersion of a velocity channel, determined by the velocity resolution ($\delta v$).}

\textbf{The impact of this artificial broadening is more significant for clouds with intrinsically narrow line widths. Here we adopt $\sigma_{\mathrm{ch}}$ = $\frac{\delta v}{\sqrt{12}}$ following \citet{Hirota2024}. When the observed dispersion is 8.5~km\,s$^{-1}$ (the median value for HVCs), the broadening contributes only $\sim1.5\%$ to the observed value. In contrast, for a dispersion of 3.8~km\,s$^{-1}$ (the median for disk clouds), the observed value can be overestimated by as much as $\sim8\%$. This implies that the true velocity dispersion of disk clouds are even smaller than observed, and thus the conclusion that HVCs have systematically larger velocity dispersions than disk clouds remains robust.}

\textbf{\textcolor{black}{Similarly, the cloud radius may also be affected by the artificial broadening due to the finite spatial resolution. However, since the distribution of radius shows no significant bias between the HVCs and disk clouds (Figuure \ref{radius_sigv} (a), \ref{Other_scaling_relation}(b)), both populations are subject to comparable degrees of broadening. Therefore, this effect does not alter our conclusion.}}

\textbf{In addition, the calculation of luminosity mass depends on the adopted $\alpha_{\rm CO}$ conversion factor. According to \citet{Lee2024}, the conversion factor in M83 is approximately half within the inner 3 kpc compared to the outer 3–5 kpc region. Since most of the HVCs are located within the central 3 kpc, their masses may be overestimated when the uniform $\alpha_{\rm CO}$ is assumed. If we adopt smaller $\alpha_{\rm CO}$ in the inner 3 kpc disk, most HVCs would shift toward the left in Figure~\ref{radius_sigv}(b), and the virial parameter would increase accordingly. 
On the other hand, when the HVCs originated from outside the galaxy and have significantly lower metallicities compared to disk clouds, their true gas masses could be underestimated, \textcolor{black}{and most HVCs would shift toward the right in the figure.}}

\section{Discussion}\label{discussion}

In this section, we discuss the origin of HVCs. Considering the lack of AGN activity in M83, two possible explanations for the origin of the HVCs are proposed. The first is the ejection by supernova explosions, and the second is an inflow from outside the galactic disk.

\begin{figure*}[htbp]
    \centering
    \includegraphics[width=0.8\textwidth]{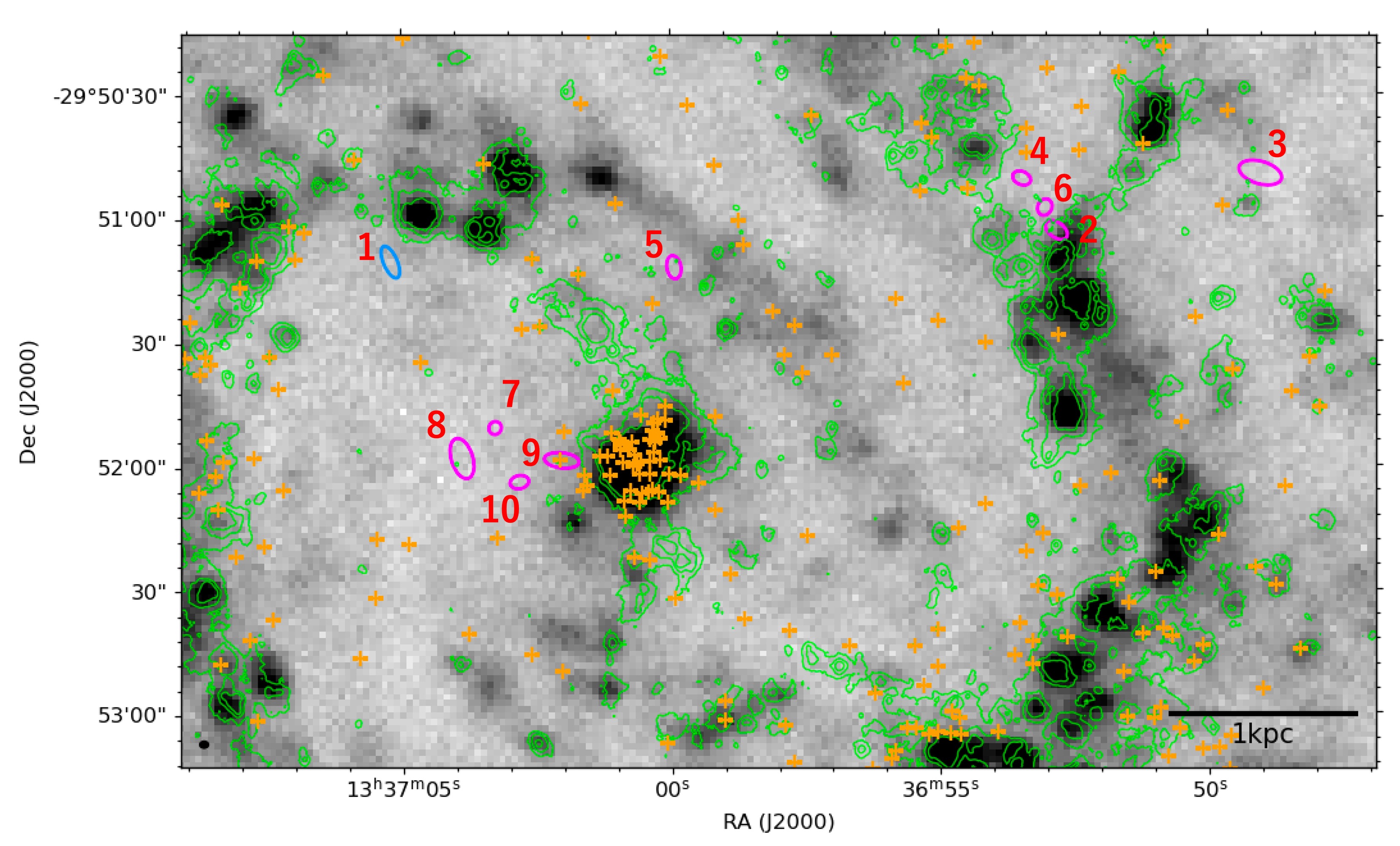}
    \caption{\textbf{\textcolor{black}{Comparison of the locations of HVCs (blue and purple ellipses) with SNR candidates from \citet{Long2022} (orange crosses). The background image shows FUV emission from GALEX \citep{GildePaz2007}, and the green contours represent H$\alpha$ emission from the SINGG survey \citep{Meurer2006}. The red numbers correspond to the HVC IDs listed in Table~\ref{table}. The sizes of the HVC ellipses are the same as those shown in Figure~\ref{number}.}}}
    \label{SNR}
\end{figure*}

First, we consider the possibility that HVCs originated from supernova explosions. Using the supernova remnant catalog from \citet{Long2022}, we plotted HVCs and supernova remnants on a single map (Figure \ref{SNR}). We found that only one HVC (ID=9) coincides with the position of a supernova remnant.
HVC 9 is likely associated with the supernova remnant, whereas the other HVCs show no correlation with supernova remnants.
\textbf{\textcolor{black}{In addition, to examine whether the HVCs are related to regions of past intense star formation that may now be producing supernova explosions, we overlaid far-ultraviolet (FUV) emission from \citet{GildePaz2007} as the background of Figure~\ref{SNR}, along with H$\alpha$ emission contours from \citet{Meurer2006}. FUV traces star formation on timescales of $\sim$100\,Myr, while H$\alpha$ traces more recent star formation within $\sim$10\,Myr. Therefore, regions bright in FUV but faint in H$\alpha$ can be interpreted as sites of active star formation at $\sim100$ Myr ago. As shown in Figure~\ref{SNR}, there is no clear spatial correlation between the HVCs and such regions of past intense star formation.
}}

\textcolor{black}{Next, to compare the cloud energy with the energy of a supernova explosion, we calculate the kinetic energy $E_{\text{kin}}$ [erg] of the molecular cloud's systemic motion using the velocity deviation $\Delta v$ between the disk velocity field and the cloud, and the molecular cloud mass $M_{\text{mol}}$, as follows:}

\begin{align}
   E_{\text{kin}} = \frac{1}{2} M_{\text{mol}} {\Delta v}^2.
\end{align}

This kinetic energy corresponds to the energy required to lift the molecular cloud from the disk if the HVC originates from it. Note that $\Delta v$ is the relative velocity along the line of sight, \textcolor{black}{\textbf{and the projection effect of their velocity vectors
has not been taken into account.}}
\textbf{\textcolor{black}{As we assume that the contribution from non-circular motion to the $\Delta v$ is negligible, the derived kinetic energy values should be considered as lower limits. Therefore, the actual energy required to eject the gas may be greater than these estimates.
}}

Regarding supernova energy, a single supernova typically releases $10^{51}$\,erg, but only $\sim10\%$ of this energy is transferred to the kinetic energy of the interstellar medium (ISM) \citep{Chevalier1974}. \textbf{\textcolor{black}{The actual efficiency is likely much lower, assuming that supernova energy is  released more or less isotropically and that only a small fraction of this energy can be directed toward the HVC.
}}
\begin{figure}[htbp]
    \centering
    \includegraphics[width=\linewidth]{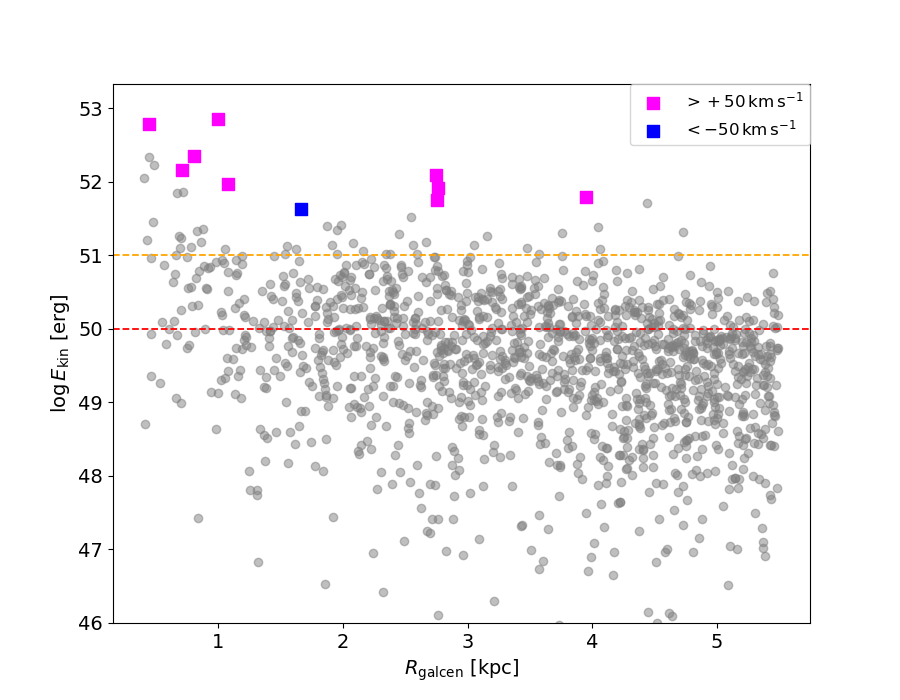}
    \caption{\textcolor{black}{The kinetic energy $E_\mathrm{kin}$ of molecular clouds against their galactocentric radii $R_\text{galcen}$: magenta squares represent molecular clouds with $\Delta v > +50\,\text{km}\,\text{s}^{-1}$, while the blue square represents a molecular cloud with $\Delta v < -50\,\text{km}\,\text{s}^{-1}$, gray dots represent disk molecular clouds. Orange dashed line shows typical supernova energy of $10^{51}$ erg and red dashed line represents $10\%$ of supernova energy expected to be transferred to the ISM \citep{Chevalier1974}.}
    }
    \label{galr_energy}
\end{figure}

Figure \ref{galr_energy} plots the kinetic energy of molecular clouds against galactocentric radius $R_\text{galcen}$. We find that the HVCs' kinetic energy exceeds $10^{51}$ erg, with some reaching up to $\sim10^{52}$ erg, and therefore, we claim that single supernova explosions are insufficient to explain their origin. \textbf{Note that galactic fountains and superbubbles, driven by multiple supernovae, can in principle reach such energies; achieving $\sim10^{52}$ erg would require on the order of 100 supernovae , \textcolor{black}{assuming an upper-limit efficiency of 10\%. However, as presented in Figure~\ref{SNR}, most HVCs show no spatial association with such intensely star-forming regions.
}}


Finally, if most HVCs originated from supernova explosions, we would expect symmetry in the number of HVCs on the redshifted and blueshifted sides of the disk. However, there is a notable excess of the HVCs on the redshifted side in Figure \ref{mom0_hvc} (b). Taking this asymmetry into account, we deduce that the origin of most HVCs is likely due to inflow from outside the galactic disk.
This conclusion is somewhat consistent with one of the scenarios proposed for feature C 
(located ~0.8 kpc east of the galactic center, spatially aligns with HVCs 7 and 8)
in \citet{Della2022}, which interprets it as non-collisional flows from multiple gas layers. 
\textbf{\textcolor{black}{They also suggested that the high-velocity component reported in CO(2--1), corresponding to HVC 8, may represent infalling material.}} However, in our study, the large velocity dispersion (Figure \ref{radius_sigv} (a)) implies a potential interaction between the HVCs and the disk component. This high dispersion might be explained by the decelerating structure observed in the MW’s inflowing gas \citep{Bieging2024}.


\citet{Lucchini2024} identified HVCs in a MW-like galaxy from the TNG50 simulations and found that most of them do not originate from the disk, but rather come from satellites or from the cooling of hot and warm CGM. Our conclusion aligns with their results, while in this study, we cannot determine specifically where our HVCs are inflowing from.


Past interactions could be a potential source of this inflow.
According to \citet{Karachentsev2005}, M83 is in a subgroup that includes 13 galaxies considered to be tidally interacting with M83. These galaxies have a projected separation of less than 380 kpc from M83 and are located at a distance of $4.5 \pm 0.5$ Mpc, while M83 itself reported at 4.47 Mpc in their study.
It has been known that in M83 the optical and kinetic nuclei are spatially offset, and \citet{Knapen2010} attributed the offset to  a past interaction with one of the subgroup members, possibly with the close neighbor galaxy NGC 5253.
\textbf{\citet{Eibensteiner2023} revealed complex large-scale H$\;${\sc i} kinematics outside the molecular disk of M83, including extended tilted rings and asymmetric velocity structures. These features may represent signatures of past minor mergers or tidal interactions.}

\textcolor{black}{Considering the lack of spatial correlation with supernova remnants, the high kinetic energy, and the distribution bias toward the redshifted side, we conclude that most HVCs in M83 are inflowing gas from outside the galactic disk, possibly due to a past interaction with a neighboring galaxy.}


\section{Summary and Conclusion }\label{conclusion}
We utilized high spatial resolution ($\sim$40 pc), high-sensitivity CO(1–0) data to search for HVCs in the nearby barred spiral galaxy M83. We identified molecular clouds using the \texttt{astrodendro} algorithm and selected those whose velocities deviate by more than 50 km s\(^{-1}\) from the disk velocity field as HVCs. As a result, we found 10 HVCs ($\sim1\%$ of all clouds) with radii of 30–80 pc, masses on the order of \(10^5\,M_{\odot}\), velocity deviations of 50–120 km s\(^{-1}\) and velocity dispersions of 3–20 km s\(^{-1}\). These HVCs exhibit a tendency toward higher velocity dispersions compared to other molecular clouds in the disk of M83,
\textbf{\textcolor{black}{suggesting that they are more likely to be gravitationally unbound.}}
Because M83 does not show any evidence of AGN activity, we considered two possible origins for these HVCs: supernovae or inflowing gas from outside the galactic disk. Given their positions relative to known supernova remnants, their kinetic energies of \(10^{51}\)–\(10^{52}\) erg, and the preferred distribution on the redshifted side of the disk, we conclude that supernovae alone are insufficient to account for their origin. Therefore, we conclude that most of these HVCs are inflowing molecular gas from outside the galactic disk. \textcolor{black}{To further constrain where the HVCs are inflowing from, H$\;${\sc i} data with high spatial resolution and sensitivity will be helpful.}

\section*{Acknowledgments}
We thank the anonymous reviewer for their constructive comments and suggestions. This paper makes use of the following ALMA data: ADS/JAO.ALMA \#2017.1.00079.S. ALMA is a partnership of ESO (representing its member states), NSF (USA) and NINS (Japan), together with NRC (Canada), NSTC and ASIAA (Taiwan), and KASI (Republic of Korea), in cooperation with the Republic of Chile. The Joint ALMA Observatory is operated by ESO, AUI/NRAO and NAOJ. The National Radio Astronomy Observatory is a facility of the National Science Foundation operated under cooperative agreement by Associated Universities, Inc..
Data analysis was carried out on the Multi-wavelength Data Analysis System operated by the Astronomy Data Center (ADC), National Astronomical Observatory of Japan.
This research made use of \texttt{astrodendro}, a Python package to compute dendrograms of Astronomical data (\url{http://www.dendrograms.org/}). This research made use of
APLpy, an open-source plotting package for Python \citep{aplpy2012,aplpy2019}.

M.N. was supported by the ALMA Japan Research Grant of NAOJ ALMA Project, NAOJ-ALMA-355.
F.M. is supported by JSPS KAKENHI grant No. JP23K13142.
J.K. acknowledges support from NSF through grants AST-2006600 and AST-2406608. F.E. was supported by JSPS KAKENHI grant No. 20H00172.
K.T. was supported by a NAOJ ALMA Scientific Research grant Nos. 2022-22B. K.K. acknowledges the support by JSPS KAKENHI Grant Numbers JP23K20035 and JP24H00004.

Facilities: ALMA

Software: CASA \citep{McMullin2007,CASA2022},
astrodendro \citep{Astrodendro2019},
Astropy \citep{Astropy},
APLpy \citep{aplpy2012, aplpy2019}, NumPy \citep{Numpy}, matplotlib \citep{matplotlib}, dendroplot \citep{dendroplot}

\appendix

\section{Position-Velocity Diagrams of HVCs}\label{appendix_PV}
\begin{figure*}[htbp]
    \centering
    \includegraphics[width=0.85\textwidth]{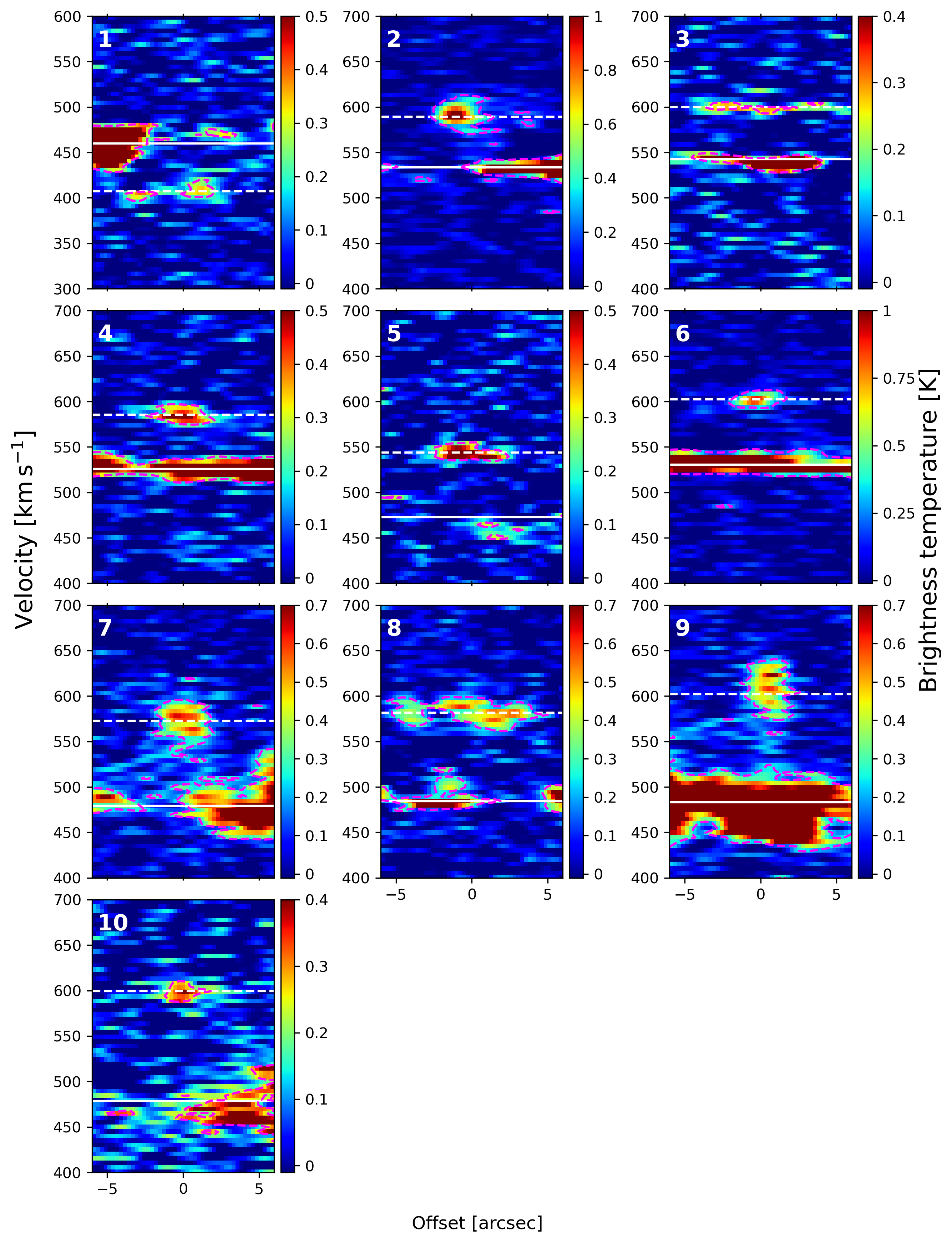}
    \caption{\textbf{Position–velocity diagrams of the HVCs, extracted along their major axes and averaged over a width of 3 pixels ($0\farcs75=15\,\rm pc$). The positive direction of Offset is chosen such that it points toward the end of the major axis that is closer to the north (i.e., higher declination). Magenta dashed contours indicate the $3\sigma$ level. The horizontal white solid line marks the disk velocity, while the horizontal white dashed line indicates $v_{\mathrm{LSR}}^{\mathrm{HVC}}$ of HVCs.}}
    \label{PV}
\end{figure*}

\textbf{Figure~\ref{PV} shows the position–velocity diagrams of the HVCs, extracted along their major axes.}

\section{Other scaling relations}\label{other_scaling}
\begin{figure*}[htbp]
    \centering
    \includegraphics[width=0.9\textwidth]{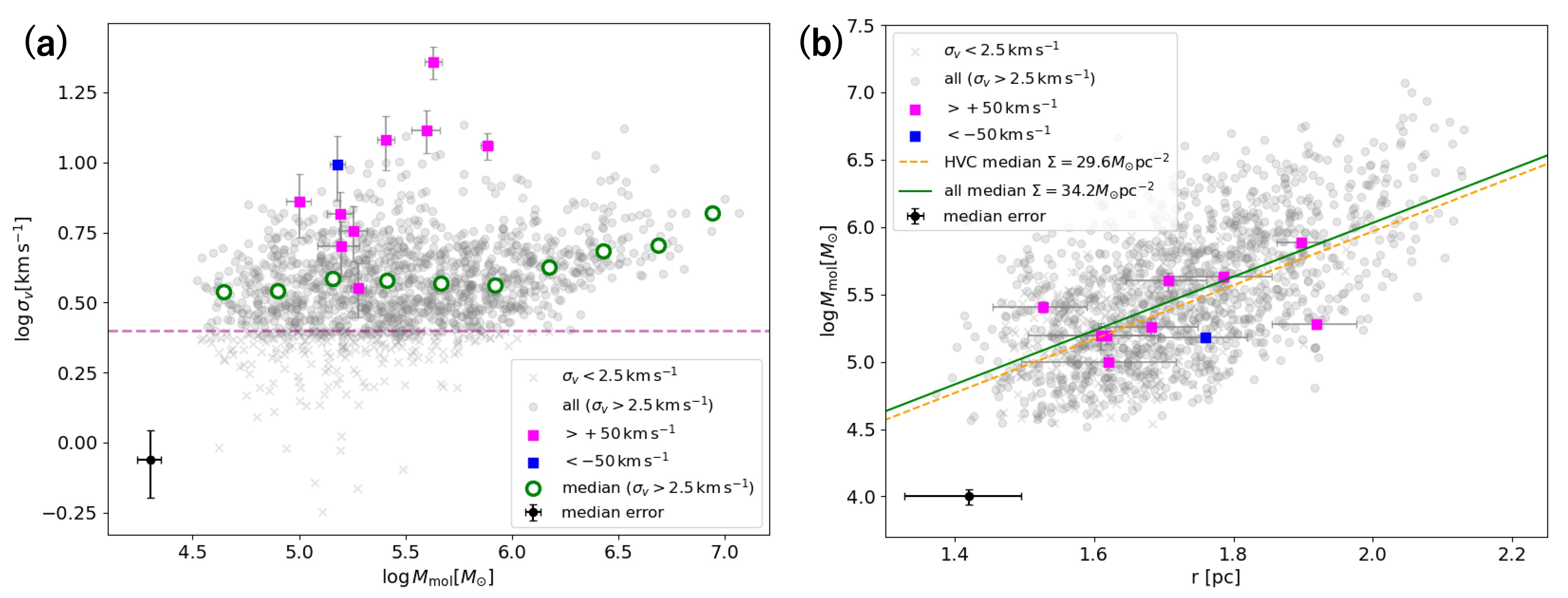}
    \caption{\textbf{Same as Figure 5 but for other scaling relations.
    (a) $M_{\rm mol}$--$\sigma_v$ relation: 
    Green dots show the median values in each bin, including HVCs but excluding clouds with $\sigma_{v} < 2.5\,\text{km}\,\text{s}^{-1}$. The purple dashed line indicates the velocity dispersion resolution limit of $2.5\,\text{km}\,\text{s}^{-1}$. }\textbf{(b) $r$--$M_{\rm mol}$ relation: 
    The green solid line indicates the median surface density $\Sigma= 34.2  M_{{\odot}} \mathrm{pc}^{-2}$ for all clouds. The orange dashed line shows the median surface density $\Sigma= 29.6  M_{{\odot}} \mathrm{pc}^{-2}$  for the HVCs.}}
    \label{Other_scaling_relation}
\end{figure*}

\textbf{Figure~\ref{Other_scaling_relation} shows (a) the $M_{\rm mol}$--$\sigma_v$ and (b) $r$--$M_{\rm mol}$ relations, respectively.  
From the $M_{\rm mol}$--$\sigma_v$ relation, we confirm—similar to the $r$--$\sigma_v$ relation in Figure~\ref{radius_sigv}(a)—that the HVCs tend to have higher velocity dispersions than the disk clouds.  
In the $r$--$M_{\rm mol}$ relation, we estimate the median surface densities ($\Sigma$) of the disk clouds and HVCs using $\Sigma = M_{\rm mol}/(\pi r^2)$, and plot them as a green solid line and orange dashed line, respectively.  
The median surface densities of the disk clouds and HVCs are approximately 30~$M_\odot$\,pc$^{-2}$ and 34~$M_\odot$\,pc$^{-2}$, respectively, indicating that the difference is not significant.
}
\clearpage

\bibliography{sample631}{}

\begin{thebibliography}{}
\expandafter\ifx\csname natexlab\endcsname\relax\def\natexlab#1{#1}\fi
\providecommand{\url}[1]{\href{#1}{#1}}
\providecommand{\dodoi}[1]{doi:~\href{http://doi.org/#1}{\nolinkurl{#1}}}
\providecommand{\doeprint}[1]{\href{http://ascl.net/#1}{\nolinkurl{http://ascl.net/#1}}}
\providecommand{\doarXiv}[1]{\href{https://arxiv.org/abs/#1}{\nolinkurl{https://arxiv.org/abs/#1}}}

\bibitem[{{Astropy Collaboration} {et~al.}(2018){Astropy Collaboration}, {Price-Whelan}, {Sip{\H{o}}cz}, {G{\"u}nther}, {Lim}, {Crawford}, {Conseil}, {Shupe}, {Craig}, {Dencheva}, {Ginsburg}, {VanderPlas}, {Bradley}, {P{\'e}rez-Su{\'a}rez}, {de Val-Borro}, {Aldcroft}, {Cruz}, {Robitaille}, {Tollerud}, {Ardelean}, {Babej}, {Bach}, {Bachetti}, {Bakanov}, {Bamford}, {Barentsen}, {Barmby}, {Baumbach}, {Berry}, {Biscani}, {Boquien}, {Bostroem}, {Bouma}, {Brammer}, {Bray}, {Breytenbach}, {Buddelmeijer}, {Burke}, {Calderone}, {Cano Rodr{\'\i}guez}, {Cara}, {Cardoso}, {Cheedella}, {Copin}, {Corrales}, {Crichton}, {D'Avella}, {Deil}, {Depagne}, {Dietrich}, {Donath}, {Droettboom}, {Earl}, {Erben}, {Fabbro}, {Ferreira}, {Finethy}, {Fox}, {Garrison}, {Gibbons}, {Goldstein}, {Gommers}, {Greco}, {Greenfield}, {Groener}, {Grollier}, {Hagen}, {Hirst}, {Homeier}, {Horton}, {Hosseinzadeh}, {Hu}, {Hunkeler}, {Ivezi{\'c}}, {Jain}, {Jenness}, {Kanarek}, {Kendrew}, {Kern}, {Kerzendorf}, {Khvalko}, {King}, {Kirkby}, {Kulkarni},
  {Kumar}, {Lee}, {Lenz}, {Littlefair}, {Ma}, {Macleod}, {Mastropietro}, {McCully}, {Montagnac}, {Morris}, {Mueller}, {Mumford}, {Muna}, {Murphy}, {Nelson}, {Nguyen}, {Ninan}, {N{\"o}the}, {Ogaz}, {Oh}, {Parejko}, {Parley}, {Pascual}, {Patil}, {Patil}, {Plunkett}, {Prochaska}, {Rastogi}, {Reddy Janga}, {Sabater}, {Sakurikar}, {Seifert}, {Sherbert}, {Sherwood-Taylor}, {Shih}, {Sick}, {Silbiger}, {Singanamalla}, {Singer}, {Sladen}, {Sooley}, {Sornarajah}, {Streicher}, {Teuben}, {Thomas}, {Tremblay}, {Turner}, {Terr{\'o}n}, {van Kerkwijk}, {de la Vega}, {Watkins}, {Weaver}, {Whitmore}, {Woillez}, {Zabalza}, \& {Astropy Contributors}}]{Astropy}
{Astropy Collaboration}, {Price-Whelan}, A.~M., {Sip{\H{o}}cz}, B.~M., {et~al.} 2018, \aj, 156, 123, \dodoi{10.3847/1538-3881/aabc4f}

\bibitem[{{Baba} {et~al.}(2016){Baba}, {Morokuma-Matsui}, {Miyamoto}, {Egusa}, \& {Kuno}}]{Baba2016}
{Baba}, J., {Morokuma-Matsui}, K., {Miyamoto}, Y., {Egusa}, F., \& {Kuno}, N. 2016, \mnras, 460, 2472, \dodoi{10.1093/mnras/stw987}

\bibitem[{{Bieging} \& {Kong}(2024)}]{Bieging2024}
{Bieging}, J.~H., \& {Kong}, S. 2024, \mnras, 531, 4138, \dodoi{10.1093/mnras/stae1419}

\bibitem[{{Blair} {et~al.}(2014){Blair}, {Chandar}, {Dopita}, {Ghavamian}, {Hammer}, {Kuntz}, {Long}, {Soria}, {Whitmore}, \& {Winkler}}]{Blair2014}
{Blair}, W.~P., {Chandar}, R., {Dopita}, M.~A., {et~al.} 2014, \apj, 788, 55, \dodoi{10.1088/0004-637X/788/1/55}

\bibitem[{{Bolatto} {et~al.}(2013){Bolatto}, {Wolfire}, \& {Leroy}}]{Bolatto2013}
{Bolatto}, A.~D., {Wolfire}, M., \& {Leroy}, A.~K. 2013, \araa, 51, 207, \dodoi{10.1146/annurev-astro-082812-140944}

\bibitem[{{Bovy} \& {Rix}(2013)}]{Bovy2013}
{Bovy}, J., \& {Rix}, H.-W. 2013, \apj, 779, 115, \dodoi{10.1088/0004-637X/779/2/115}

\bibitem[{{CASA Team} {et~al.}(2022){CASA Team}, {Bean}, {Bhatnagar}, {Castro}, {Donovan Meyer}, {Emonts}, {Garcia}, {Garwood}, {Golap}, {Gonzalez Villalba}, {Harris}, {Hayashi}, {Hoskins}, {Hsieh}, {Jagannathan}, {Kawasaki}, {Keimpema}, {Kettenis}, {Lopez}, {Marvil}, {Masters}, {McNichols}, {Mehringer}, {Miel}, {Moellenbrock}, {Montesino}, {Nakazato}, {Ott}, {Petry}, {Pokorny}, {Raba}, {Rau}, {Schiebel}, {Schweighart}, {Sekhar}, {Shimada}, {Small}, {Steeb}, {Sugimoto}, {Suoranta}, {Tsutsumi}, {van Bemmel}, {Verkouter}, {Wells}, {Xiong}, {Szomoru}, {Griffith}, {Glendenning}, \& {Kern}}]{CASA2022}
{CASA Team}, {Bean}, B., {Bhatnagar}, S., {et~al.} 2022, \pasp, 134, 114501, \dodoi{10.1088/1538-3873/ac9642}

\bibitem[{{Chevalier}(1974)}]{Chevalier1974}
{Chevalier}, R.~A. 1974, \apj, 188, 501, \dodoi{10.1086/152740}

\bibitem[{{Chomiuk} \& {Povich}(2011)}]{Chomiuk2011}
{Chomiuk}, L., \& {Povich}, M.~S. 2011, \aj, 142, 197, \dodoi{10.1088/0004-6256/142/6/197}

\bibitem[{{Della Bruna} {et~al.}(2022){Della Bruna}, {Adamo}, {Amram}, {Rosolowsky}, {Usher}, {Sirressi}, {Schruba}, {Emsellem}, {Leroy}, {Bik}, {Blair}, {McLeod}, {{\"O}stlin}, {Renaud}, {Robert}, {Rousseau-Nepton}, \& {Smith}}]{Della2022}
{Della Bruna}, L., {Adamo}, A., {Amram}, P., {et~al.} 2022, \aap, 660, A77, \dodoi{10.1051/0004-6361/202142315}

\bibitem[{{Dessauges-Zavadsky} {et~al.}(2007){Dessauges-Zavadsky}, {Combes}, \& {Pfenniger}}]{Dessauges2007}
{Dessauges-Zavadsky}, M., {Combes}, F., \& {Pfenniger}, D. 2007, \aap, 473, 863, \dodoi{10.1051/0004-6361:20077277}

\bibitem[{{Di Teodoro} {et~al.}(2020){Di Teodoro}, {McClure-Griffiths}, {Lockman}, \& {Armillotta}}]{DiTeodoro2020}
{Di Teodoro}, E.~M., {McClure-Griffiths}, N.~M., {Lockman}, F.~J., \& {Armillotta}, L. 2020, \nat, 584, 364, \dodoi{10.1038/s41586-020-2595-z}

\bibitem[{{Eibensteiner} {et~al.}(2023){Eibensteiner}, {Bigiel}, {Leroy}, {Koch}, {Rosolowsky}, {Schinnerer}, {Sardone}, {Meidt}, {de Blok}, {Thilker}, {Pisano}, {Ott}, {Barnes}, {Querejeta}, {Emsellem}, {Puschnig}, {Utomo}, {Be{\v{s}}li{\'c}}, {den Brok}, {Faridani}, {Glover}, {Grasha}, {Hassani}, {Henshaw}, {Jim{\'e}nez-Donaire}, {Kerp}, {Dale}, {Kruijssen}, {Laudage}, {Sanchez-Blazquez}, {Smith}, {Stuber}, {Pessa}, {Watkins}, {Williams}, \& {Winkel}}]{Eibensteiner2023}
{Eibensteiner}, C., {Bigiel}, F., {Leroy}, A.~K., {et~al.} 2023, \aap, 675, A37, \dodoi{10.1051/0004-6361/202245290}

\bibitem[{{Elmegreen} {et~al.}(1998){Elmegreen}, {Chromey}, \& {Warren}}]{Elmegreen1998}
{Elmegreen}, D.~M., {Chromey}, F.~R., \& {Warren}, A.~R. 1998, \aj, 116, 2834, \dodoi{10.1086/300657}

\bibitem[{{Fukui} {et~al.}(2021){Fukui}, {Habe}, {Inoue}, {Enokiya}, \& {Tachihara}}]{Fukui2021}
{Fukui}, Y., {Habe}, A., {Inoue}, T., {Enokiya}, R., \& {Tachihara}, K. 2021, \pasj, 73, S1, \dodoi{10.1093/pasj/psaa103}

\bibitem[{{Gil de Paz} {et~al.}(2007){Gil de Paz}, {Boissier}, {Madore}, {Seibert}, {Joe}, {Boselli}, {Wyder}, {Thilker}, {Bianchi}, {Rey}, {Rich}, {Barlow}, {Conrow}, {Forster}, {Friedman}, {Martin}, {Morrissey}, {Neff}, {Schiminovich}, {Small}, {Donas}, {Heckman}, {Lee}, {Milliard}, {Szalay}, \& {Yi}}]{GildePaz2007}
{Gil de Paz}, A., {Boissier}, S., {Madore}, B.~F., {et~al.} 2007, \apjs, 173, 185, \dodoi{10.1086/516636}

\bibitem[{{Gillmon} {et~al.}(2006){Gillmon}, {Shull}, {Tumlinson}, \& {Danforth}}]{Gillmon2006}
{Gillmon}, K., {Shull}, J.~M., {Tumlinson}, J., \& {Danforth}, C. 2006, \apj, 636, 891, \dodoi{10.1086/498053}

\bibitem[{Harris {et~al.}(2020)Harris, Millman, van~der Walt, Gommers, Virtanen, Cournapeau, Wieser, Taylor, Berg, Smith, Kern, Picus, Hoyer, van Kerkwijk, Brett, Haldane, del Río, Wiebe, Peterson, Gérard-Marchant, Sheppard, Reddy, Weckesser, Abbasi, Gohlke, \& Oliphant}]{Numpy}
Harris, C.~R., Millman, K.~J., van~der Walt, S.~J., {et~al.} 2020, Nature, 585, 357–362, \dodoi{10.1038/s41586-020-2649-2}

\bibitem[{Hirota {et~al.}(2014)Hirota, Kuno, Baba, Egusa, Habe, Muraoka, Tanaka, Nakanishi, \& Kawabe}]{Hirota2014}
Hirota, A., Kuno, N., Baba, J., {et~al.} 2014, Publications of the Astronomical Society of Japan, 66, \dodoi{10.1093/pasj/psu006}

\bibitem[{{Hirota} {et~al.}(2024){Hirota}, {Koda}, {Egusa}, {Sawada}, {Sakamoto}, {Heyer}, {Lee}, {Maeda}, {Boissier}, {Calzetti}, {Elmegreen}, {Harada}, {Ho}, {Kobayashi}, {Kuno}, {Madore}, {Mart{\'\i}n}, {Donovan Meyer}, {Muraoka}, \& {Watanabe}}]{Hirota2024}
{Hirota}, A., {Koda}, J., {Egusa}, F., {et~al.} 2024, \apj, 976, 198, \dodoi{10.3847/1538-4357/ad8228}

\bibitem[{Hunter(2007)}]{matplotlib}
Hunter, J.~D. 2007, Computing in Science \& Engineering, 9, 90, \dodoi{10.1109/MCSE.2007.55}

\bibitem[{{Jarrett} {et~al.}(2019){Jarrett}, {Cluver}, {Brown}, {Dale}, {Tsai}, \& {Masci}}]{Jarrett2019}
{Jarrett}, T.~H., {Cluver}, M.~E., {Brown}, M.~J.~I., {et~al.} 2019, \apjs, 245, 25, \dodoi{10.3847/1538-4365/ab521a}

\bibitem[{Karachentsev(2005)}]{Karachentsev2005}
Karachentsev, I.~D. 2005, The Astronomical Journal, 129, 178–188, \dodoi{10.1086/426368}

\bibitem[{{Knapen} {et~al.}(2010){Knapen}, {Sharp}, {Ryder}, {Falc{\'o}n-Barroso}, {Fathi}, \& {Guti{\'e}rrez}}]{Knapen2010}
{Knapen}, J.~H., {Sharp}, R.~G., {Ryder}, S.~D., {et~al.} 2010, \mnras, 408, 797, \dodoi{10.1111/j.1365-2966.2010.17180.x}

\bibitem[{{Koda} {et~al.}(2023){Koda}, {Hirota}, {Egusa}, {Sakamoto}, {Sawada}, {Heyer}, {Baba}, {Boissier}, {Calzetti}, {Donovan Meyer}, {Elmegreen}, {Gil de Paz}, {Harada}, {Ho}, {Kobayashi}, {Kuno}, {Lee}, {Madore}, {Maeda}, {Mart{\'\i}n}, {Muraoka}, {Nakanishi}, {Onodera}, {Pineda}, {Scoville}, \& {Watanabe}}]{Koda2023}
{Koda}, J., {Hirota}, A., {Egusa}, F., {et~al.} 2023, \apj, 949, 108, \dodoi{10.3847/1538-4357/acc65e}

\bibitem[{{Larson}(1969)}]{Larson1969}
{Larson}, R.~B. 1969, \mnras, 145, 271, \dodoi{10.1093/mnras/145.3.271}

\bibitem[{{Lee} {et~al.}(2024){Lee}, {Koda}, {Hirota}, {Egusa}, \& {Heyer}}]{Lee2024}
{Lee}, A.~M., {Koda}, J., {Hirota}, A., {Egusa}, F., \& {Heyer}, M. 2024, \apj, 968, 97, \dodoi{10.3847/1538-4357/ad40a0}

\bibitem[{{Long} {et~al.}(2022){Long}, {Blair}, {Winkler}, {Della Bruna}, {Adamo}, {McLeod}, \& {Amram}}]{Long2022}
{Long}, K.~S., {Blair}, W.~P., {Winkler}, P.~F., {et~al.} 2022, \apj, 929, 144, \dodoi{10.3847/1538-4357/ac5aa3}

\bibitem[{{Lucchini} {et~al.}(2024){Lucchini}, {Han}, {Hernquist}, \& {Conroy}}]{Lucchini2024}
{Lucchini}, S., {Han}, J.~J., {Hernquist}, L., \& {Conroy}, C. 2024, \apj, 974, 105, \dodoi{10.3847/1538-4357/ad6dde}

\bibitem[{Mac~Low \& Klessen(2004)}]{Mac_Low2004}
Mac~Low, M.-M., \& Klessen, R.~S. 2004, Reviews of Modern Physics, 76, 125–194, \dodoi{10.1103/revmodphys.76.125}

\bibitem[{{McKee} \& {Zweibel}(1992)}]{Mckee1992}
{McKee}, C.~F., \& {Zweibel}, E.~G. 1992, \apj, 399, 551, \dodoi{10.1086/171946}

\bibitem[{{McMullin} {et~al.}(2007){McMullin}, {Waters}, {Schiebel}, {Young}, \& {Golap}}]{McMullin2007}
{McMullin}, J.~P., {Waters}, B., {Schiebel}, D., {Young}, W., \& {Golap}, K. 2007, in Astronomical Society of the Pacific Conference Series, Vol. 376, Astronomical Data Analysis Software and Systems XVI, ed. R.~A. {Shaw}, F.~{Hill}, \& D.~J. {Bell}, 127

\bibitem[{{Meurer} {et~al.}(2006){Meurer}, {Hanish}, {Ferguson}, {Knezek}, {Kilborn}, {Putman}, {Smith}, {Koribalski}, {Meyer}, {Oey}, {Ryan-Weber}, {Zwaan}, {Heckman}, {Kennicutt}, {Lee}, {Webster}, {Bland-Hawthorn}, {Dopita}, {Freeman}, {Doyle}, {Drinkwater}, {Staveley-Smith}, \& {Werk}}]{Meurer2006}
{Meurer}, G.~R., {Hanish}, D.~J., {Ferguson}, H.~C., {et~al.} 2006, \apjs, 165, 307, \dodoi{10.1086/504685}

\bibitem[{{Miller} {et~al.}(2009){Miller}, {Bregman}, \& {Wakker}}]{Miller2009}
{Miller}, E.~D., {Bregman}, J.~N., \& {Wakker}, B.~P. 2009, \apj, 692, 470, \dodoi{10.1088/0004-637X/692/1/470}

\bibitem[{{Miville-Desch{\^e}nes} {et~al.}(2005){Miville-Desch{\^e}nes}, {Boulanger}, {Reach}, \& {Noriega-Crespo}}]{Miville2005}
{Miville-Desch{\^e}nes}, M.~A., {Boulanger}, F., {Reach}, W.~T., \& {Noriega-Crespo}, A. 2005, \apjl, 631, L57, \dodoi{10.1086/496961}

\bibitem[{{Putman} {et~al.}(2012){Putman}, {Peek}, \& {Joung}}]{Putman2012}
{Putman}, M.~E., {Peek}, J.~E.~G., \& {Joung}, M.~R. 2012, \araa, 50, 491, \dodoi{10.1146/annurev-astro-081811-125612}

\bibitem[{Putman {et~al.}(2011)Putman, Saul, \& Mets}]{Putman_2011}
Putman, M.~E., Saul, D.~R., \& Mets, E. 2011, Monthly Notices of the Royal Astronomical Society, 418, 1575–1586, \dodoi{10.1111/j.1365-2966.2011.19524.x}

\bibitem[{Robitaille(2019)}]{aplpy2019}
Robitaille, T. 2019, {APLpy v2.0: The Astronomical Plotting Library in Python}, \dodoi{10.5281/zenodo.2567476}

\bibitem[{{Robitaille} \& {Bressert}(2012)}]{aplpy2012}
{Robitaille}, T., \& {Bressert}, E. 2012, {APLpy: Astronomical Plotting Library in Python}, Astrophysics Source Code Library.
\newblock \doeprint{1208.017}

\bibitem[{{Robitaille} {et~al.}(2019){Robitaille}, {Rice}, {Beaumont}, {Ginsburg}, {MacDonald}, \& {Rosolowsky}}]{Astrodendro2019}
{Robitaille}, T., {Rice}, T., {Beaumont}, C., {et~al.} 2019, {astrodendro: Astronomical data dendrogram creator}, Astrophysics Source Code Library, record ascl:1907.016

\bibitem[{{Rosolowsky} \& {Leroy}(2006)}]{Rosolowsky2006}
{Rosolowsky}, E., \& {Leroy}, A. 2006, \pasp, 118, 590, \dodoi{10.1086/502982}

\bibitem[{Sancisi {et~al.}(2008)Sancisi, Fraternali, Oosterloo, \& van~der Hulst}]{Sancisi2008}
Sancisi, R., Fraternali, F., Oosterloo, T., \& van~der Hulst, T. 2008, The Astronomy and Astrophysics Review, 15, 189–223, \dodoi{10.1007/s00159-008-0010-0}

\bibitem[{{Solomon} {et~al.}(1987){Solomon}, {Rivolo}, {Barrett}, \& {Yahil}}]{Solomon1987}
{Solomon}, P.~M., {Rivolo}, A.~R., {Barrett}, J., \& {Yahil}, A. 1987, \apj, 319, 730, \dodoi{10.1086/165493}

\bibitem[{{Thim} {et~al.}(2003){Thim}, {Tammann}, {Saha}, {Dolphin}, {Sandage}, {Tolstoy}, \& {Labhardt}}]{Thim2003}
{Thim}, F., {Tammann}, G.~A., {Saha}, A., {et~al.} 2003, \apj, 590, 256, \dodoi{10.1086/374888}

\bibitem[{{Veilleux} {et~al.}(2005){Veilleux}, {Cecil}, \& {Bland-Hawthorn}}]{Veilleux2005}
{Veilleux}, S., {Cecil}, G., \& {Bland-Hawthorn}, J. 2005, \araa, 43, 769, \dodoi{10.1146/annurev.astro.43.072103.150610}

\bibitem[{{Wakker}(1991)}]{Wakker1991}
{Wakker}, B.~P. 1991, \aap, 250, 499

\bibitem[{{Wakker}(2006)}]{Wakker2006}
---. 2006, \apjs, 163, 282, \dodoi{10.1086/500365}

\bibitem[{Wong(2020)}]{dendroplot}
Wong, T. 2020, {dendroplot}, \url{https://github.com/tonywong94/dendroplot}

\bibitem[{{Wong} {et~al.}(2017){Wong}, {Hughes}, {Tokuda}, {Indebetouw}, {Bernard}, {Onishi}, {Wojciechowski}, {Bandurski}, {Kawamura}, {Roman-Duval}, {Cao}, {Chen}, {Chu}, {Cui}, {Fukui}, {Montier}, {Muller}, {Ott}, {Paradis}, {Pineda}, {Rosolowsky}, \& {Sewi{\l}o}}]{Wong2017}
{Wong}, T., {Hughes}, A., {Tokuda}, K., {et~al.} 2017, \apj, 850, 139, \dodoi{10.3847/1538-4357/aa9333}

\end{thebibliography}
\bibliographystyle{aasjournal}



\end{document}